\documentclass[%
reprint,
showpacs,preprintnumbers,
 amsmath,amssymb,
 aps,
 prd,
]{revtex4-1}

\usepackage{graphicx}

\begin{document}

\preprint{Imperial/TP/10/DJW/01}	
\title{Studying a relativistic field theory at finite chemical potential with the density matrix renormalization group}
\author{David J. Weir}
\email{david.weir03@imperial.ac.uk}
\affiliation{%
Theoretical Physics, Blackett Laboratory, Imperial College London, SW7 2AZ London, United Kingdom.
}

\date{\today} 
\begin{abstract}
The density matrix renormalization group is applied to a relativistic complex scalar field at finite chemical potential. The two-point function and various bulk quantities are studied. It is seen that bulk quantities do not change with the chemical potential until it is larger than the minimum excitation energy. The technical limitations of the density matrix renormalization group for treating bosons in relativistic field theories are discussed. Applications to other relativistic models and to nontopological solitons are also suggested.
\end{abstract}
\maketitle

\section{Introduction}
In relativistic field theory on the Euclidean lattice, the chemical potential manifests itself as an imaginary vector potential in the imaginary time direction~\cite{Hasenfratz:1983ba}. The use of Monte Carlo techniques and reliance on importance sampling has the effect of making the probability weight meaningless if the action or the fermion determinant is complex; this is known as the ``sign problem''. With lattice fermions, the sign problem is also a consequence of a complex fermion determinant resulting from the inclusion of a chemical potential term. A variety of techniques are commonly used to circumvent this problem, including Glasgow reweighting~\cite{Barbour:1997ej} and analytic continuation to an imaginary chemical potential~\cite{Roberge:1986mm}.

The aim of this paper is to use the density matrix renormalization group (DMRG) to study a toy model in relativistic quantum field theory with a nonzero chemical potential. The results of this paper demonstrate that this approach correctly captures the phenomenology of the theory. We seek the ground state of the Hamiltonian for the system using variational methods and so DMRG does not suffer from the aforementioned sign problem~\cite{Sugihara:2005cf}. This problem of relativistic field theories at finite density should be contrasted with the negative sign problem of fermions in quantum Monte Carlo~\cite{Dagotto:1994zz}. It is already widely acknowledged that this is avoided by DMRG~\cite{Hallberg:2006ch,PhysRevLett.81.445}. As far as we are aware, however, no attempt has been made to circumvent the relativistic finite-density sign problem with DMRG.

When the chemical potential $\mu$ is smaller than the lowest excitation energy (at zero coupling, the bare mass $m$), the bulk quantities studied are seen to have the same value as at $\mu=0$. This effect is known from QCD as the ``Silver Blaze'' problem~\cite{Aarts:2008wh}. Such an effect cannot be studied in a nonrelativistic field theory.

For larger chemical potentials, the system condenses. We cannot voyage far into the condensed phase because the truncation in states converges more slowly. For smaller values of the chemical potential, however, convergence is better. Nonetheless, the truncation in bosonic states and the limitation to $(1+1)$ dimensions remain the main drawbacks of using DMRG as a nonperturbative tool in quantum field theory.

The density matrix renormalization group has previously been applied to bosonic problems with nonzero chemical potential for condensed-matter systems~\cite{Urba:2006}. In such cases the models usually describe a single bosonic field and there is no symmetry between the particles and antiparticles. Such Hamiltonians emerge as the nonrelativistic limit of the model considered here when the parts of the Hamiltonian concerning the antiparticles and their interactions are neglected~\cite{Evans:1995yz}. The current work, however, accurately describes the fully relativistic $\mathrm{U}(1)$ model. Toy models in particle physics have been studied using DMRG previously, in particular a one-component scalar field with $\lambda\phi^4$ interaction~\cite{Sugihara:2004qr}, the massive Schwinger model~\cite{Byrnes:2002nv} and a simple model exhibiting asymptotic freedom~\cite{MartinDelgado:1999jb}. $\mathrm{SU}(N)$ spin chains have also been studied~\cite{Fuhringer08}.

In this paper we consider a scalar field model with $\mathrm{U}(1)$ symmetry in $(1+1)$ dimensions. For convenience, we transform fields to a two-component real scalar field, giving a Lagrangian
\begin{equation}
 \mathcal{L} = \frac{1}{2}(\partial_\mu \phi_n)(\partial^\mu \phi_n) - \frac{1}{2}m^2 \phi_n \phi_n - \frac{1}{4}\lambda (\phi_n \phi_n)^2.
\end{equation}
The conjugate momenta are $\pi_n = \dot{\phi}_n$. We then perform a Legendre transform to give the corresponding Hamiltonian density for this theory~\cite{Haber:1981ts},
\begin{equation}
 \mathcal{H}_0 = \frac{1}{2}(\nabla \phi_n)(\nabla \phi_n) + \frac{1}{2} \pi_n\pi_n + \frac{1}{2} m^2 \phi_n \phi_n + \frac{1}{4}\lambda(\phi_n \phi_n)^2.
\end{equation}
This model has one conserved charge
\begin{equation}
\label{eq:charge}
 Q=\int \mathrm{d} x \; j_0 = \int \mathrm{d} x \; \left[\phi_1 \pi_2 -\phi_2  \pi_1 \right].
\end{equation}
We introduce the chemical potential $\mu$ as a Lagrange multiplier into an effective Hamiltonian $\mathcal{H}$ for minimizing the energy at nonzero total charge $Q$ and obtain
\begin{equation}
\label{eq:effectiveham}
 \begin{split}
 \mathcal{H} & =  \mathcal{H}_0 - \mu j_0 \\
  & =  \frac{1}{2}(\nabla \phi_n)(\nabla \phi_n) + \frac{1}{2} \pi_n\pi_n + \frac{1}{2} m^2 \phi_n \phi_n + \frac{1}{4} \lambda (\phi_n \phi_n)^2 \\
	& \quad - \mu \left( \phi_1\pi_2  - \phi_2 \pi_1  \right),
 \end{split}
\end{equation}
which we will use for our studies of this model using the DMRG.

This paper is organized as follows. In Sec. \ref{sec:dmrg} we outline the density matrix renormalization group as applied to the present model and discuss places where our implementation differs from those in the literature. Our numerical results are presented in Sec. \ref{sec:results}, and our final remarks can be found in Sec. \ref{sec:conclusion}.

\section{The density matrix renormalization group}
\label{sec:dmrg}

The density matrix renormalization group is a variational technique for finding quantum states of quasi-one dimensional systems~\cite{White93}. Originally conceived to study systems -- such as the Heisenberg model -- on lattices too large to treat with exact diagonalization, it is essentially a development of Wilson's numerical renormalization group for handling interacting systems~\cite{Schollwock05,Noack,Hallberg:2006ch}. A brief qualitative summary is given here. The entire system is termed the ``superblock,'' and is divided into two renormalized ``system blocks'', between which one or more sites are ``inserted''. This allows numerical diagonalization of a smaller Hamiltonian than if the entire system were exactly diagonalized. Typically, one is only interested in the ground state, and so only one eigenvector needs to be numerically obtained. Operators are transformed to best represent the states of interest and the inserted site is incorporated into one of the system blocks. The process repeats with a new ``inserted site''.

When this technique is used to study a finite system as in the present case, the representation of the system is optimized by repeated ``sweeping'' to one end of the lattice and then the other. The system size is kept constant by growing one system block at the expense of the other. Sweeping stops when observables of interest no longer change. If the system is homogeneous then the symmetries of the system can be exploited to accelerate this process.

Our approach is to insert only a single site, rather than two sites, during the sweeping process. This approach has been used successfully to study the one-component $\lambda\phi^4$ model in Ref.~\cite{Sugihara:2004qr}, and its use remains appropriate here. Generally, when fermionic systems are considered, the number of states per lattice site required to fully describe the system is not large and indeed for small lattices it would in principle be possible to exactly diagonalize the system. In addition, there are unambiguous reasons why adding two sites helps to improve efficiency when using the infinite-volume algorithm with spin systems~\cite{White93}, including that it keeps the system symmetric. However, in our present system we must truncate the tower of bosonic states and even for a small system there is no way to exactly diagonalize the Hamiltonian. As each inserted site gets incorporated into our renormalized blocks the highest state accessible is given by the highest state in the truncation used for the single site. 

Let us label the number of states kept in the system block by $M$ and the number of states on the inserted site $N$. By the above argument it seems reasonable to expect that, for bosonic systems with a given maximum dimension $D$ of the Hamiltonian, the best numerical results will be obtained by taking the largest possible truncation for three sites $(D \approx M^2 N)$ rather than for four $(D \approx M^2 N^2)$. 

Unlike in Monte Carlo simulations, where the preference is for periodic or twisted boundary conditions, for DMRG calculations the results are more precise when obtained with open boundary conditions. The interaction between the extreme ends of the lattice cannot be renormalized within the traditional scheme for finite system DMRG and hence the system only has an approximate translational invariance. A translationally invariant approach is possible when one reformulates the problem in terms of matrix-product states~\cite{Verstraete:2004zza,Verstraete:2010ft}. Treating bosons will remain difficult, due to the high bond dimension of the matrices needed.  For the purposes of this paper, it is sufficient to consider the `traditional' choice of boson number states, with open boundary conditions.

\subsection{The model}

A single-component relativistic scalar field was first studied with the DMRG in Refs.~\cite{Sugihara:2004qr,Sugihara:2004xp}. This discussion therefore parallels these works, extended to a two-component field. Moving to a $(L\times\infty)$ lattice, we keep the conjugate momenta as operators $\pi_n$ but discretize the gradient term in (\ref{eq:effectiveham}) in the usual way,
\begin{equation}
 \nabla \phi_n \to \frac{1}{a} \left[\phi_n(x+a) - \phi_n(x)\right],
\end{equation}
where we have introduced a lattice spacing $a$ and label the sites of the lattice by $0,\dotsc,x,x+a,\dotsc,L$. We can set $a=1$ without loss of generality as we can vary $m$, $\lambda$ and $\mu$ instead. We work with real-space DMRG (although momentum-space formulations exist), which means we need a set of basis states for every lattice site. We treat $\phi_1$ and $\phi_2$ as separate fields with their own creation and annihilation operators,
\begin{equation}
\label{eq:quantize1}
 \phi_n(x) = \frac{1}{\sqrt{2}} \left(a_n^\dag(x) + a_n(x)\right),
\end{equation}
$n=1,2$, and
\begin{equation}
\label{eq:quantize2}
\pi_n(x) = \frac{i}{\sqrt{2}} \left(a_n^\dag(x) - a_n(x)\right)
\end{equation}
where $a_n^\dag$ and $a_n$ create and annihilate particles of type $n$ at sites labeled by $x$ on the lattice. This motivates the use of $\left| p, q \right>$ for boson number states at a given lattice site, where $p$ and $q$ label each of the two different particle types.

The equal time commutation relation,
\begin{equation}
\label{eq:etcr}
 \left[ \phi_n(x), \pi_m(y) \right] = i\delta_{x,y}\delta_{m,n},
\end{equation}
becomes
\begin{equation}
 \left[ a_n(x), a_m^\dag(y) \right] = \delta_{x,y}\delta_{m,n}.
\end{equation}
The Hamiltonian becomes
\begin{equation}
\begin{split}
 H_0 & =  \sum_{x=1}^L \left(\frac{1}{2}\left[ \pi_1(x)^2 +\pi_2(x)^2 \right] \right. \\ & \qquad + \; \frac{1}{2} m^2 \left[\phi_1(x)^2 + \phi_2(x)^2 \right]  \\ & \qquad + \;\left.   \frac{1}{4} \lambda \left[\phi_1(x)^2 + \phi_2(x)^2 \right]^2 \right) \\
&\quad+ \sum_{x=1}^{L-1} \frac{1}{2} \left(\left[\phi_1(x) - \phi_1(x+1)\right]^2 \right. \\ & \qquad + \; \left. \left[\phi_2(x) - \phi_2(x+1)\right]^2 \right).
\end{split}
\end{equation}
We then arrive at the effective Hamiltonian on the lattice
\begin{equation}
 H = H_0 - \sum_{x=1}^L \mu \left[\pi_2 (x) \phi_1(x) - \pi_1(x) \phi_2(x)\right].
\end{equation}
Quantized as outlined above, $H_0$ can always be written as a real symmetric matrix. A nonzero chemical potential requires us to diagonalize a complex, Hermitian Hamiltonian $H$  but this has been done previously in DMRG studies of electron systems with persistent currents~\cite{Meden:2003oq}. It does not present any problem for the DMRG.

We split the Hamiltonian up into the Hamiltonians for the two renormalized `system' blocks and the single inserted site, plus interaction terms~\cite{Sugihara:2004qr},
\begin{equation}
 H = H_L + h_{n-1,n} + h_n + h_{n,n+1} + H_R
\end{equation}
where $H_L$ and $H_R$ are the left and right system blocks respectively. The single site $h_n$ is
\begin{multline}
\label{eq:cham}
 h_n = \frac{1}{2}\left[ \pi_1(n)^2 +\pi_2(n)^2 \right]+ \frac{1}{2} m^2 \left[\phi_1(n)^2 + \phi_2(n)^2 \right] \\
 + \frac{1}{4} \lambda \left[\phi_1(n)^2 + \phi_2(n)^2 \right]^2 + \mu \left[\pi_2 (n) \phi_1(n) - \pi_1(n) \phi_2(n)\right],
\end{multline}
and the interaction term $h_{n,n+1}$ is given by
\begin{equation}
 h_{n,n+1} = \frac{1}{2} \left(\left[\phi_1(n) - \phi_1(n+1)\right]^2 + \left[\phi_2(n) - \phi_2(n+1)\right]^2 \right).
\end{equation}
Diagonalizing $H$ numerically, we obtain an approximation to the ground state $\left| \psi \right>$,
\begin{equation}
\label{eq:state}
 \left| \psi \right> = \sum_{i,j,k} \psi_{ijk} \left| i \right> \left| j \right> \left| k \right>,
\end{equation}
where the single labels $i$, $j$ and $k$ run over the truncated bases. The left and right blocks are in general renormalized (except at the ends of the lattice) and the basis states are optimized, but the central site's basis $\left| j \right>$ always corresponds to a sum over the two-particle basis states $\left| p, q \right>$.

The matrix elements are stated for $\lambda\phi^4$ with a single scalar field in Ref.~\cite{Sugihara:2004xp}, and have been summarized in a form generalized to the present case in Appendix \ref{sec:matrixelements}.

Regardless of what numerical technique we use to study the discretized Hamiltonian system, we will have to truncate the basis states in (\ref{eq:state}). This can be interpreted as a UV cutoff that may not always be as high as that of the lattice spacing, $1/a$.

\subsection{Relation with microscopic Hamiltonians for Bose-Einstein condensation}
The density matrix renormalization group has been used in condensed-matter systems to study microscopic Hamiltonians that exhibit Bose-Einstein condensation, such as Bose-Hubbard models~\cite{Urba:2006,Rapsch99}. These models are substantially different from the present case as there is no possibility of spontaneous symmetry breaking; there is only a single particle species present. In this section we show how the present work corresponds to relativistic physics that cannot be obtained with a microscopic Hamiltonian motivated by problems in condensed matter.

For the current discussion, let us work with free fields ($\lambda=0$). In condensed-matter systems, Bose-Einstein condensates are often modeled by the Gross-Pitaevskii equation. Following Ref.~\cite{Evans:1995yz}, we therefore want to transform to nonrelativistic fields $\Psi$ and $\bar{\Psi}$ that (together with their complex conjugates) satisfy the Gross-Pitaevskii equations of motion with a nonrelativistic chemical potential $\mu_{\text{nr}}$,
\begin{equation}
 \left(-i\frac{\partial}{\partial t} + \frac{k^2}{2m^2} - \mu_{\text{nr}}\right)\Psi = 0,
\end{equation}
and equivalently for $\bar{\Psi}$. In terms of $\phi_1$ and $\phi_2$ (and their conjugate momenta), the appropriate transformation is
\begin{eqnarray}
\phi_1(x) + i\; \phi_2(x) & = & \frac{1}{\sqrt{2\omega}}\left( \Psi(x) + \bar{\Psi}^*(x) \right), \label{eq:trans1} \\
\pi_1(x) + i\;\pi_2(x) & = & i\sqrt{\frac{\omega}{2}}\left( \Psi^*(x) - \bar{\Psi}(x)  \right), \label{eq:trans2}
\end{eqnarray}
where $\omega = \sqrt{k^2 + m^2}$. One should then expand to order $k^2$. If we compare the free-field action for the relativistic fields
\begin{equation}
 S_{\text{rel}} = \int dt \int dx \sum_{n=1}^2 \; \phi_n\left(\frac{\partial^2}{\partial t^2} - \frac{\partial^2}{\partial x^2} + m^2\right)\phi_n,
\end{equation}
with that for one of the nonrelativistic fields
\begin{equation}
 S_{\text{nr}} = \int dt \int dx \; \Psi^{*} \left[i \frac{\partial}{\partial t} + \frac{1}{2m}\nabla^2 + \mu - m\right]\Psi
\end{equation}
then we see that the nonrelativistic fields split into two parts, one for the $\Psi$ field and one for the $\bar{\Psi}$ field. We have omitted the nonrelativistic fields $\bar{\Psi}$ and $\bar{\Psi}^*$ corresponding to one of the particle species, as they are suppressed by a factor $e^{-(\mu+m)\bar{\Psi}^{*}\bar{\Psi}}$ in the partition function. The transformations (\ref{eq:trans1}-\ref{eq:trans2}) have had the added effect of diagonalizing the chemical potential term for each particle species -- the term in the Hamiltonian referring to the chemical potential is now just $\mu\; \Psi^{*}\Psi$, a number operator that is usually combined with $m$ to give $\mu_{\text{nr}}$. Hence, the sign problem disappears for nonrelativistic models at finite density as one can treat the system with one field.

\subsection{Extension to finite temperature}
\label{sec:temp}
Ideally, we would be able to study nonzero density at \textsl{finite} temperature. Since we keep the conjugate momenta $\pi_n$ as operators throughout we are dealing with a $(1+1)$-dimensional system, and cannot introduce finite-temperature physics in the standard, Euclidean, way. For the DMRG, the traditional method used to treat small $T>0$ is to find several low-lying states~\cite{Hallberg:2006ch,Moukouri96}
\begin{equation}
|\psi^{(n)}\rangle = \psi^{(n)}_{ijk}| i \rangle | j \rangle |
 k \rangle
\end{equation}
when numerically diagonalizing the Hamiltonian, and weight them by the Boltzmann factor in the reduced density matrix $\rho$ to give
\begin{equation}
\label{eq:finitetdm}
 \rho = \sum_n e^{-\beta E_n} \sum_{k'} \psi^{(n)}_{ijk'}\psi^{*\;(n)}_{i'j'k'}.
\end{equation}
This adds an extra layer of truncation: we must make sure that the number of excited states included in the density matrix is large enough to correctly capture the finite-temperature behavior to the accuracy allowed by our truncation. The most straightforward way to implement this is to vary the number of eigenvectors obtained by our numerical diagonalization code for inclusion in Eq. \ref{eq:finitetdm}. If increasing the number of eigenvectors obtained does not affect the observables of interest then we can consider the truncation adequate. This ought to happen when the smallest obtained Boltzmann weight, associated with the highest energy, $e^{-\beta E_{n_\text{max}}}$ is comparable in size to our algorithm's convergence tolerance.

These added complications mean that finite-temperature behavior for a bosonic system cannot be reliably obtained by the traditional method, given present computing capacity. A more promising direction is to use the transfer-matrix DMRG (TDMRG) method to handle a discretized imaginary time direction of finite size~\cite{Wang:1997}. This should give good results at high temperature, and permits access to thermodynamic quantities (but not long-distance correlation functions). However, one would then have to abandon the na\"{i}ve chemical potential term used in the present work and add an imaginary, constant vector potential as is usual in Euclidean lattice studies~\cite{Hasenfratz:1983ba}.

\subsection{Measurements}

\label{sec:correl}

For expectation values of observables defined on a single lattice site it is desirable to take measurements at the center of the lattice, using the single inserted site. Then for an observable $\mathcal{O}$ (for example $j_0$) defined on the single site, we have
\begin{equation}
 \left< \mathcal{O} \right> = \sum_{ijj'k} \psi^{(0)*}_{ijk} [\mathcal{O}(x)]_{jj'}\psi^{(0)}_{ij'k} = \mathrm{Tr} \,\rho \mathcal{O}.\\
\end{equation}

We calculate the correlation length $\xi$ in the system by looking at the decay of the two-point function. This serves as a useful cross-check to verify that the onset of a nonzero particle density occurs, as one would expect, close to $\mu=m$ at weak coupling. Unlike in Monte Carlo simulations, we do not have access to anything other than equal time correlations.

To calculate the two-point function
\begin{equation}
 C_{ab}(x-y)=\left< \phi_a(x) \phi_b(y) \right> - \left<\phi_a(x)\right>\left<\phi_b(y)\right>,
\end{equation}
we must store the operators for $\phi_a(x)$ and $\phi_b(y)$ until we are ready to calculate the two-point function
\begin{equation}
 \left< \phi_a(x) \phi_b(y) \right>  = \sum_{ii'jkk'} \psi^{(0)*}_{ijk} [\phi_a(x)]_{ii'} [\phi_b(y)]_{kk'} \psi^{(0)}_{i'jk'} \\
\end{equation}
and disconnected pieces
\begin{eqnarray}
 \left< \phi_a(x) \right> & = & \sum_{ii'jk} \psi^{(0)*}_{ijk} [\phi_a(x)]_{ii'}\psi^{(0)}_{i'jk} \\
 \left< \phi_b(y) \right> & = & \sum_{ijkk'} \psi^{(0)*}_{ijk} [\phi_b(y)]_{kk'}\psi^{(0)}_{ijk'}.
\end{eqnarray}
So that the sum is over a complete set of states, it is simplest from a practical point of view if the two operators come from different DMRG blocks~\cite{Noack}. Since our main interest in the two-point function is to determine the renormalized mass from the long-distance behavior, there is no reason to choose operators on the same block. Therefore we sacrifice measurements of the short-distance behavior to give better results at longer distances.

\section{Results and computational considerations}
\label{sec:results}

The renormalized block Hamiltonians $H$ are Hermitian matrices of size $M^2 N \times M^2 N$. The main limitation on the accuracy and quality of the calculations presented here is the memory required to store these matrices (even when represented in a sparse format, this is over a gigabyte), and to a much lesser extent the additional computation time required by $\mu \neq 0$. One ``sweep'' to the right or the left in the algorithm takes several hours of computer time for the largest truncations used here, although this can be less depending on the value of $\mu$; for $\mu=0$ the Lanczos algorithm is used and performance is much better. It also depends on the number of eigenvectors required, making $T\neq 0$ prohibitively expensive given current computational resources.

Computations were carried out with exhaustive searches of the parameter space. Numerical diagonalization routines from NAG were used; the superblock was stored as a sparse matrix. For $\mu=0$, the Hamiltonian is a real, symmetric matrix and the Lanczos method can be used. The current work, however, is primarily concerned with $\mu\neq 0$, for which it is easiest to use an implicitly restarted Arnoldi algorithm. For all the results shown, we tolerate a fractional error of at most $10^{-8}$ in the numerically obtained Ritz vectors.

The ground state energy is obtained directly when diagonalizing the Hamiltonian. Other measurements are made as described in Sec. \ref{sec:correl}. We wait until results change by no more than $1$ part in $10^6$ between sweeps before considering a quantity adequately converged.

One naturally expects the charge operator $Q$ to be self-adjoint so that in adding a term $-\mu Q$ to form an effective Hamiltonian we must have $\mu$ real. Unlike in calculations where the chemical potential is analytically continued, we have not made any changes to the original model. With enough memory and computer time we could work with $\mu \gg m^2$ , $M$ and $N$ very large and study nonzero particle densities far from the onset of condensation. There is no sign problem, merely the problem of representing a condensate using the DMRG. However, it is accepted that the picture given by a finite truncation serves to capture the physics~\cite{Hallberg:2006ch}.

\subsection{Convergence with truncation size and arrangement}
\label{sec:convergence}

\begin{figure}[t]
\centering
\includegraphics[scale=0.333,angle=270]{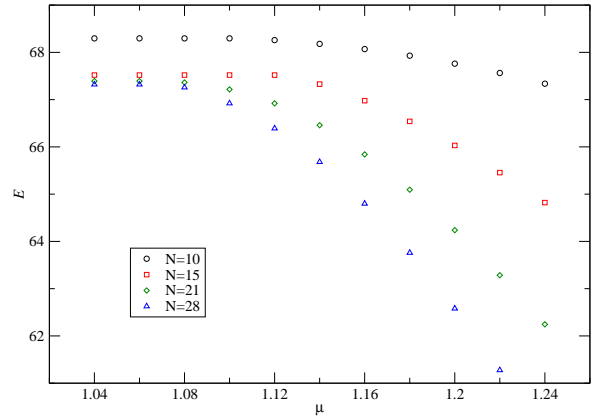}
\caption{\label{fig:twostyles} Truncating the tower of bosonic states, with $L=40$, $m^2=1$, $\lambda=0.1$ and $M=N$. The ground state energy is shown as a function of $\mu$ for various consistently truncated bases.}
\end{figure}

Noting that storage considerations are a limiting factor, we must consider the most reasonable approximate basis for the lattice sites. For simplicity, let us take the number of sites in the renormalized blocks $M$ to equal those of the single site $N$. Finite-truncation effects are more noticeable when varying $M$, the renormalized block size.

We choose a consistent way of truncating the bosonic states at each site as, in principle, there is an infinite tower that must be truncated. One might choose to fix either the maximum number of bosons of \textsl{any} type $\{ \left|i,j\right> \big| i\leq n, j\leq n\}$, or the maximum number of bosons of \textsl{either} type $\{ \left|i,j\right> \big| i + j \leq n \}$. It seems better to adopt the latter organization, as this is the truncation that better respects the global symmetry of the theory. Consider a field transformation
\begin{equation}
\begin{split}
\phi_1 & \to \phi_1\cos(\theta) + \phi_2\sin(\theta); \\
\phi_2 & \to \phi_1\cos(\theta) - \phi_2\sin(\theta).
\end{split}
\end{equation}
To quantize this equivalent field theory, we can define new creation and annihilation operators in terms of the old ones that will obey the same canonical commutation relations
\begin{equation}
\begin{split}
a^\dag_1(x) & \to a^\dag_1(x)\cos(\theta) + a^\dag_2(x)\sin(\theta); \\
a^\dag_2(x) & \to a^\dag_1(x)\cos(\theta) - a^\dag_2(x)\sin(\theta).
\end{split}
\end{equation}
Therefore if we use the truncated basis $\{ \left|i,j\right> \big| i + j \leq n \}$, then we will always be able to reexpress the rotated state using the truncated basis whereas starting with the truncated basis $\{ \left|i,j\right> \big| i\leq n, j\leq n\}$ risks leaving us outside the span. Given this choice, Figure \ref{fig:twostyles} shows how varying the number of states in the renormalized blocks affects the ground state energy.

\begin{figure}[t]
\centering
\includegraphics[scale=0.333,angle=270]{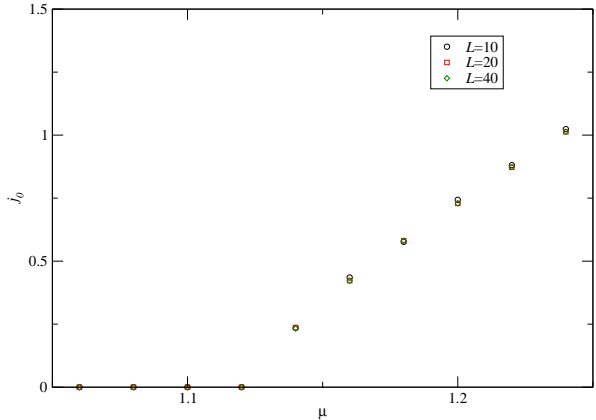}
\caption{\label{fig:finitevol} Plot of the charge density $j_0$ as a function of $\mu$ at various $L$, for $m^2=1$, $\lambda=0.1$, and $M=N=28$. There is no discernible finite-size effect for the spatial lattice size. It should be emphasized that for $\mu$ smaller than that shown here the charge density is exactly zero down to $\mu=0$.}
\end{figure}

It is clear that, for a bosonic system, the DMRG converges quickly for zero particle density. It can also be reliably used to determine the onset of nonzero particle density. However, for large particle densities the truncation is overwhelmed and accuracy is greatly diminished.

\subsection{Condensation and finite-size effects}

\begin{figure}[t]
\centering
\includegraphics[scale=0.333,angle=270]{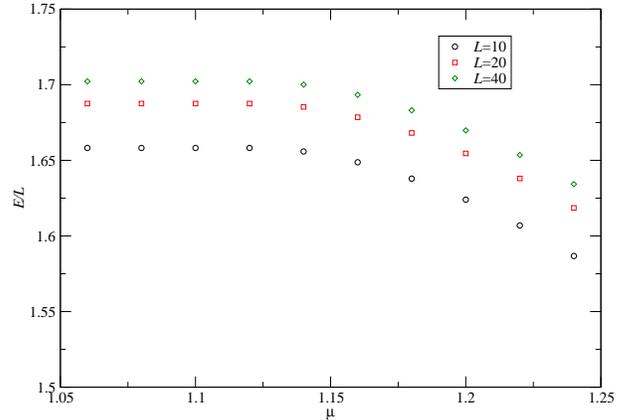}
\caption{\label{fig:finitevolen} Plot showing the ground state energy per site $E/L$ as a function of $\mu$, for $m^2=1$, $\lambda=0.1$, and $M=N=28$ at various $L$. The finite-size effect is much more prominent here than in Fig. \ref{fig:finitevol}.}
\end{figure}

For $\lambda=0.1$ and various lattice sizes $L$, the particle density $j_0$ measured at the center of the lattice is shown in Fig. \ref{fig:finitevol}. These show a transition to a nonzero particle density as $\mu$ is increased, with no discernible dependence on $L$. On the other hand, the ground state energy per site -- shown in Fig. \ref{fig:finitevolen} -- has a very clear dependence on volume. The volume dependence is qualitatively similar for all values of the chemical potential studied.

In Ref.~\cite{Sugihara:2004qr} it was noticed that more sweeps were required for convergence close to the critical coupling. For $\mu$ close to condensation, a similar effect is observed.

\subsection{Two-point function and phenomenological accuracy}

\begin{figure}[t]
\centering

\includegraphics[scale=0.333,angle=270]{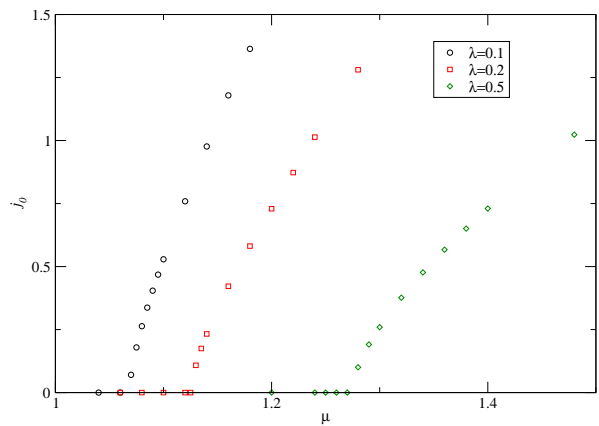}
\caption{\label{fig:chempot} The charge density $j_0$ is plotted as a function of $\mu$, for $m^2=1$, $L=40$, and $M=N=28$ at various $\lambda$. The phenomenologically expected zero charge density is observed for $\mu<\mu_C$.}
\end{figure}

\begin{figure}[t]
\centering

\includegraphics[scale=0.333,angle=270]{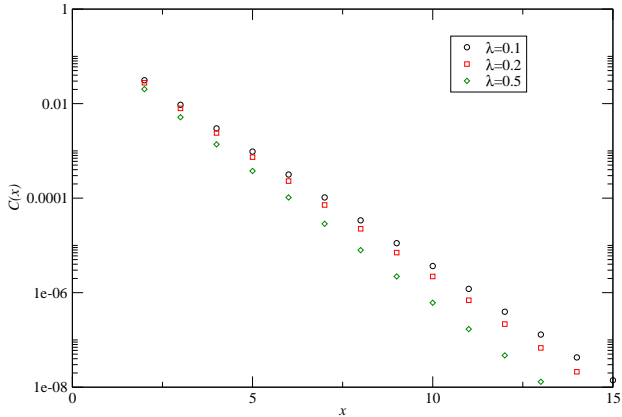}
\caption{\label{fig:twopoint} Example measurements of the two-point correlator $C(x)$ for various $\lambda$, $m^2=1$ with volume $L=40$ and truncation $M=N=28$.}
\end{figure}

As $\lambda$ is increased from zero, the chemical potential at which the particle density becomes nonzero moves away from the free-field result. Figure \ref{fig:chempot} shows this for $L=40$. We would expect that the condensation occurs when $\mu \gtrsim m_R$, the renormalized mass, at weak coupling. The behavior of the two-point function is shown in Figure \ref{fig:twopoint} for various $\lambda$. We obtain the scalar masses nonperturbatively from these results by a fit to the long-distance behavior $C(x) = A e^{-m x}/\sqrt{x}$. As is familiar from studies in lattice Monte Carlo, the ``plateau'' in the fitting errors is taken, to eliminate undesirable short-distance behavior (typically, the plateau is found at $x\approx 6$). Finite-size effects were determined to be negligible by comparison with similar measurements taken with $L=10$ and $L=20$. Similarly, there was no discernible difference with results for a smaller truncation on the same lattice volume ($L=40$ and $N=M=21$). The measurements from these calculations agree within errors with the results summarized in Table \ref{tab:onset}.

Taken together, these plots demonstrate that the present technique sidesteps the sign problem, and only gives a nonzero particle density in the phenomenologically expected region $\mu \gtrsim m_R$. As shown in Table \ref{tab:onset}, we can separately estimate the value of the chemical potential at which we see nonzero particle density from the data shown in Figure \ref{fig:chempot}, and the renormalized mass from fits to data comparable to that of Fig. \ref{fig:twopoint}. Taking the onset of nonzero particle density as occurring halfway between the last parameter choice where $\left<j_0\right>=0$ and the first for which $\left<j_0\right>\neq 0$, the numbers are consistent for small $\lambda$.

\begin{table}[b]
\begin{tabular}{|r|l|l|}
\hline
$\lambda$ & $m_\mathrm{R}$ & $\mu_{\text{con}}$ \\
\hline
0.1 & $1.071 \pm 0.001$ & $1.065 \pm 0.005$ \\
0.2 & $1.114 \pm 0.001$ & $1.123 \pm 0.003$ \\
0.5 & $1.231 \pm 0.001$ & $1.275 \pm 0.005$ \\
\hline
\end{tabular}
\centering
\caption{\label{tab:onset} Comparing renormalized masses and values of the chemical potential at which condensation occurs. For $m_R$, the error quoted is the estimated error in the plateaued fit whereas for $\mu_{\text{con}}$ it is half the spacing between parameter choices.}
\end{table}

\section{Conclusions}
\label{sec:conclusion}

We have shown that it is possible to use the DMRG to study bosonic, relativistic quantum field theories with two components in $(1+1)$ dimensions at zero temperature but nonzero chemical potential. The method correctly captures the formation of a nonzero particle density for large chemical potentials. It is our hope that it could serve as a useful numerical bridge from work in condensed-matter systems using microscopic Hamiltonians to studies of relativistic field theories at finite density.

Our discussion has considered various sources of error, and how best to capture the nonperturbative physics with a finite truncation in boson number states. With improvements in numerical algorithms, and by taking advantage of developments originally employed in condensed-matter contexts, the DMRG shows promise as an alternative to other numerical methods in relativistic field theory. With better computing resources, the results described here can be extended to finite temperature through the techniques described in Sec.~\ref{sec:temp}.

The applicability of the DMRG to bosonic systems will always be limited by the truncations and approximations involved: to a finite basis per site, to a finite volume, and to machine precision or worse in convergence of the Arnoldi algorithm. Fermionic systems do not suffer from the first of these three issues. Hence, as a step towards the long-term goal of studying QCD, one could revisit the massive Schwinger or Thirring models with nonzero chemical potential and at finite temperature~\cite{AlvarezEstrada:1997ja}.

Another innate limitation of DMRG is to one spatial dimension. Fortunately, much progress has been made by approaching the problem from a different direction. By identifying suitable tensor network states and corresponding renormalization procedures, systems in higher spatial dimensions can be studied. Such approaches include projected entangled pair states~\cite{Verstraete:2004cf} and the multiscale entanglement renormalization ansatz~\cite{PhysRevLett.99.220405,Vidal:2008zz}. With these methods, the major issue to overcome is the large bond dimension needed to treat the system discussed in the present work; this will make adequate system sizes computationally too expensive for the time being~\cite{Cirac:2009zz}.

The work in this paper has been carried out in the grand canonical ensemble. We can work in the canonical ensemble by finding low-lying eigenvectors of the total charge $Q$. This compels us to simultaneously diagonalize $H$ and $Q$. If we have a potential with interactions that allow a stable Q-ball to form~\cite{Coleman:1985ki,Tsumagari:2008bv}, then we would anticipate that such an inhomogeneity would appear. This would allow the nonperturbative, relativistic study of nontopological solitons, even if the possibilities are limited by the lack of dynamics and the restrictive dimensionality.

Perhaps one can also go the other way, from particle physics to condensed matter. A single-component Bose-Hubbard model can be used to study nontopological solitons in nonrelativistic field theory; these may be thought of as Q-ball analogs~\cite{Enqvist:2003zb}. With such a simple model, it is possible to explicitly work in a sector of fixed charge by choice of basis~\cite{White93}, and then one simply diagonalizes the Hamiltonian as usual to find the ground state and excitations of the nontopological soliton. The interactions necessary to create such an object have not, to our knowledge, been studied with DMRG in either condensed matter or particle physics.

\acknowledgments

We are grateful to A.~O.~Parry for introducing us to the DMRG technique. We thank T.~S.~Evans, A.~Rajantie and R.~J.~Rivers for insightful discussions. This work was supported by the Science and Technology Facilities Council. We have made use of the Imperial College High Performance Computing Service.

\appendix

\section{Matrix elements}
\label{sec:matrixelements}

Here we give the matrix elements for a position-space discretized two-component scalar field with an $O(2)$ symmetry, an extension of the work in Ref.~\cite{Sugihara:2004qr}.  We begin with the operators for the individual fields and their conjugate momenta. On a given lattice site with basis states $\left| m, n \right>$ it is straightforward to see, given (\ref{eq:quantize1}-\ref{eq:quantize2}) and requiring normalization,
\begin{widetext}
\begin{eqnarray}
 \left< m,n \right|\phi_1\left| m',n'\right> & = & \frac{1}{\sqrt{2}} \left[\sqrt{m-1}\delta_{m-1,m'} + \sqrt{m'-1}\delta_{m,m'-1}\right]\delta_{n,n'}, \\
 \left< m,n \right|\phi_2\left| m',n'\right> & = & \frac{1}{\sqrt{2}} \left[\sqrt{n-1}\delta_{n-1,n'} + \sqrt{n'-1}\delta_{n,n'-1}\right]\delta_{m,m'}, \\
 \left< m,n \right|\pi_1\left| m',n'\right> & = & \frac{i}{\sqrt{2}} \left[\sqrt{m-1}\delta_{m-1,m'} - \sqrt{m'-1}\delta_{m,m'-1}\right]\delta_{n,n'}\; \text{ and}\\
 \left< m,n \right|\pi_2\left| m',n'\right> & = & \frac{i}{\sqrt{2}} \left[\sqrt{n-1}\delta_{n-1,n'} - \sqrt{n'-1}\delta_{n,n'-1}\right]\delta_{m,m'}.
\end{eqnarray}
To construct higher moments of the fields, we must write explicitly
\begin{eqnarray}
\left< m,n \right| \pi_a\pi_a \left| m',n'\right>  & = &  \left<m,n\right|\pi_1^2  + \pi_2^2 \left| m',n' \right>, \\ 
\left< m,n \right| \phi_a\phi_a \left| m',n'\right> & = & \left< m,n \right| \phi_1^2 + \phi_2^2 \left| m',n'\right>\;\text{ and} \\
\left< m,n \right| (\phi_a\phi_a)^2 \left| m',n'\right> & = & \left< m,n \right| \phi_1^4 + 2\phi_1^2\phi_2^2 +  \phi_2^4 \left| m',n'\right>.
\end{eqnarray}
For the momenta components we then have
\begin{equation}
\left<m,n\right|\pi_1^2 \left| m',n' \right> = \frac{1}{2} \left[ - \sqrt{(m-1)(m-2)}\delta_{m-1,m'+1} +  (2m-1)\delta_{m,m'} - \sqrt{(m'-1)(m'-2)}\delta_{m+1,m'-1}\right]\delta_{n,n'}
\end{equation}
and
\begin{equation}
 \left<m,n\right|\pi_2^2 \left| m',n' \right> = \frac{1}{2} \left[ - \sqrt{(n-1)(n-2)}\delta_{n-1,n'+1} + \;(2n-1)\delta_{n,n'} - \sqrt{(n'-1)(n'-2)}\delta_{n+1,n'-1}\right]\delta_{m,m'}.
\end{equation}
Finally, for the field components we have
\begin{eqnarray}
 \left< m,n \right| \phi_1^2 \left| m',n'\right> & = & \frac{1}{2}\left[\sqrt{(m-1)(m-2)}\delta_{m-1,m'+1}  + (2 m -1)\delta_{m,m'} + \sqrt{(m'-1)(m'-2)}\delta_{m+1,m'-1}\right]\delta_{n,n'}, \\
 \left< m,n \right| \phi_2^2 \left| m',n'\right> & = & \frac{1}{2}\left[\sqrt{(n-1)(n-2)}\delta_{n-1,n'+1}  + (2 n -1)\delta_{n,n'} + \sqrt{(n'-1)(n'-2)}\delta_{n+1,n'-1}\right]\delta_{m,m'},\\
 \left< m,n \right| \phi_1^4 \left| m',n'\right> & = & \begin{aligned}[t] &
\frac{1}{4}\left[\sqrt{(m-1)(m-2)(m-3)(m-4)} \delta_{m-1,m'+3} \right. \\
 & + 4m\sqrt{(m-1)(m-2)}\delta_{m,m'+2} - 6 \sqrt{(m-1)(m-2)}\delta_{m-1,m'+1}  \\
 & - 6\sqrt{(m'-1)(m'-2)}\delta_{m+1,m'-1} + 3(2m^2 - 2m +1)\delta_{m,m'} + 4m'\sqrt{(m'-1)(m'-2)}\delta_{m+2,m'} \\
 & \left. + \sqrt{(m'-1)(m'-2)(m'-3)(m'-4)}\delta_{m+3,m'-1} \right]\delta_{n,n'},\end{aligned} \\
\left< m,n \right| \phi_2^4 \left| m',n'\right> & = & \begin{aligned}[t] &
\frac{1}{4}\left[\sqrt{(n-1)(n-2)(n-3)(n-4)} \delta_{n-1,n'+3} \right. \\ 
 & + 4n\sqrt{(n-1)(n-2)}\delta_{n,n'+2} - 6 \sqrt{(n-1)(n-2)}\delta_{n-1,n'+1} \\
 & - 6\sqrt{(n'-1)(n'-2)}\delta_{n+1,n'-1} + 3(2n^2 - 2n +1)\delta_{n,n'} + 4n'\sqrt{(n'-1)(n'-2)}\delta_{n+2,n'} \\
 & \left. + \sqrt{(n'-1)(n'-2)(n'-3)(n'-4)}\delta_{n+3,n'-1} \right]\delta_{m,m'}.\end{aligned}
\end{eqnarray}
\end{widetext}

\bibliography{dmrg}

\end{document}